\documentclass[preprint]{aastex}
\usepackage{geometry}
\usepackage[dvipdfm, pdfstartview=FitH]{hyperref}
\hypersetup{colorlinks=true,linkcolor=red,citecolor=red,filecolor=blue,urlcolor=blue}

\newcommand{\cg}{\color{green}}

\begin{document}

\title{\huge{The Science and Technology of FAST}}
\author{\textit{Sci. China-Phys. Mech. Astron.}, May 2019 Vol. 62 No. 5: 959501}

\author{}
\begin{large}

The mutual development of wireless communications and radio astronomy dates back to the 1930s when Karl Guthe Jansky discovered the electrical disturbances at the center of the Galaxy using his merry-go-round. After more than 80 years of development, radio astrophysics has fully matured, with significant contributions in both technology (receiver, electronics, and mechanical engineering) as well as science (origin and evolution of the universe, extreme events, and fundamental laws). In other words, we would not understand the universe so completely without radio astronomy.

While radio astronomy played an important role in past advancements of our understanding, even better insights are expected in the future. The physical processes associated with extreme environments occur in the universe, and radio telescopes with increasingly higher degrees of sensitivity are becoming necessary to explain many cosmic mysteries. The FAST (Five-hundred-meter Aperture Spherical radio Telescope), which is the largest single-dish radio telescope in the world, will not only promote scientific and technological discoveries in China but will also be beneficial to international scientists. Construction of FAST has been completed recently and test observations have been performed. In this special issue, we present a comprehensive review of FAST, from the dish to the receiver and the data systems; in addition, several scientific achievements obtained by FAST have also been discussed. The special issue is the first general introduction of the operational capabilities of FAST to all scientists and engineers, both domestic and international.

This special-topic issue of \textit{Science Chin-Physics, Mechanics \& Astronomy} includes seven papers. In the first invited review,~\cite{jian19} provides an exhaustive overview of the commissioning progress of FAST, and this physical examination-like report provides interesting insight to help scientists who are considering the proposal of observational programs. It is well known that pulsar and neutral hydrogen are two most important FAST science projects, and thus, the remaining six papers focus on these endeavors. Thanks to FAST¡¯s high sensitivity, it is now possible to observe rotating radio transients in greater detail~\citep{lu19a}. Single-pulse studies for pulsars over an ultra-wideband range could also be superior to those associated with other telescopes. While~\cite{yu19} demonstrates observations of the interesting abnormal emission-shift event,~\cite{lu19b} analyzes the sub-pulses of a bright radio pulsar. A drift scan survey, the supposed CRAFTS, has been proposed to simultaneously observe neutral hydrogen and pulsars. CRAFTS HI-galaxies are extensively discussed in an unprecedented way~\citep{zhan19}, and a new ensemble classification system for CRAFTS pulsar candidates has also been proposed~\citep{wang19}. Finally, in the form of a letter, ~\cite{qian19} describes the first pulsar discovered by FAST.

We would like to express our sincere gratitude to all the authors who have contributed to this special issue and our appreciation to the referees who conducted an extensive
review of the submissions, which was instrumental in narrowing down our list of contributions to the seven exceptional articles in this issue.

\begin{flushright}
Peng Jiang, Bo Peng, Di Li\\
National Astronomical Observatories

Ren-Xin Xu\\
Peking University
\end{flushright}

\end{large}

\clearpage

\end{document}